\documentstyle{article}

\begin{document}
\newcommand{\beq}{\begin{equation}}
\newcommand{\eeq}{\end{equation}}
\newcommand{\bea}{\begin{eqnarray}}
\newcommand{\eea}{\end{eqnarray}}
\newcommand{\bal}{\begin{array}{ll}} \newcommand{\eal}{\end{array}}
\newcommand{\nn}{\nonumber}
\newcommand{\hl}{{\hat \lambda}}
\newcommand{\cm}{{\cal M}}
\def\Sup{\mathop{\rm\Sup}\nolimits}
\def\R {\rm R \kern -.35cm I \kern .19cm}
\def\C{ {\rm C \kern -.15cm \vrule width.5pt \kern .12cm}}
\def\Z{ {\rm Z \kern -.30cm \angle \kern .02cm}}
\def\1{ {\rm 1 \kern -.10cm I \kern .14cm}}
\def\N{ {\rm N \kern -.31cm I \kern .15cm}}

\begin{titlepage}
\begin{flushright}
LPTHE Orsay 95/18 \\SPhT Saclay T95/042 \\May 1995
\end{flushright}
\vskip 1cm
\centerline{\bf {DYNAMICAL MASS MATRICES FROM EFFECTIVE SUPERSTRING
THEORIES\footnote{Supported
in part by the CEC Science no SC1-CT91-0729.}}} \vskip 1.5cm
\centerline{\bf{Pierre BINETRUY$^*$ , Emilian DUDAS$^{**}$}}
\vskip .5cm
\centerline{$^*$ Laboratoire de Physique Th\'eorique et Hautes Energies
\footnote{Laboratoire
associ\'e au CNRS-URA-D0063.}}
\centerline{Universit\'e Paris-Sud, B\^atiment 211, 91405 ORSAY CEDEX, FRANCE}
\vskip .5cm
\centerline{$^{**}$ Service de Physique Th\'eorique de Saclay}
\centerline{91191 GIF SUR YVETTE CEDEX, FRANCE}
\vskip 1.5cm
\begin{abstract}
We analyze the general structure of the fermion mass matrices in
effective superstrings. They are generically given at low energy by
non-trivial functions of the gauge singlet moduli fields. Interesting
structures appear in particular if they are homogeneous functions of
zero degree in the moduli. In this case we find Yukawa  matrices very
similar to the ones obtained by imposing a $U(1)$ family symmetry
to reproduce the observed hierarchy of masses and mixing angles.
The role of the $U(1)$ symmetry is played here by the modular symmetry.
Explicit orbifold examples are given where realistic quark mass matrices
can be obtained. Finally, a complete scenario is proposed which
generates the observed hierarchies in a dynamical way. \end{abstract}
\end{titlepage}

\section{Introduction.}
One of the outstanding problems of the Standard Model and its
extensions is to understand the fermion masses and mixings. These
are usually arbitrary parameters and the large mass hierarchy
observed experimentally is still to be understood. Indeed, based
on naive naturality arguments, the corresponding Yukawa
couplings are expected to be all of order one . The explanation of the
observed hierarchy certainly calls for new physics beyond the Standard
Model.

The solutions to this problem proposed to this date fall essentially
into two categories. One is a symmetry approach, first emphasized by
Froggatt and Nielsen \cite{FN}, which has been largely studied in the
literature. It postulates a new abelian horizontal gauge symmetry
spontaneously broken at a high energy scale $M_X$. The 3
families of quarks and leptons have different charges under the
corresponding $U(1)_X$ group so that only a small number of the
Standard Model Yukawa interactions be allowed by the symmetry $U(1)_X$.
All the others appear through non-renormalisable couplings to a
field whose vacuum expectation value $<\phi >$ breaks the horizontal
symmetry. In the effective theory below the scale of breaking, this
typically yields Yukawa couplings of the form
\begin{equation}
\lambda_{ij} = \left( {<\phi > \over M_X} \right)^{n_{ij}},
\label{eq:X}
\end{equation}
where $n_{ij}$ depends on the $U(1)_X$ charges of the relevant fields.
If $\varepsilon \equiv <\!\phi\!>~/~M_X$ is a small parameter, the
hierarchy of masses and mixing angles is easily obtained by assigning
different charges for different fermions.

A second approach, of dynamical origin, was recently proposed
\cite{N,Z,KPZ,BDu,BD}. The main idea is to treat Yukawa couplings as
dynamical variables to be fixed by the minimization of the vacuum energy
density. In this case, one can show that a large hierarchy can be
naturally obtained provided that the Yukawa couplings are subject to
constraints. Such a constraint could be obtained by an ad hoc imposition
of the absence of quadratic divergences in the vacuum energy
\cite{N,BDu} or as an approximate infrared evolution of the
renormalization group equations \cite{KPZ}. In ref. \cite{BD}, a
geometric origin for these constraints was proposed, related to the
properties of the moduli space in effective superstring theories. In
order to illustrate the idea, we give a simple example of a model
containing two moduli fields $T_1$, $T_2$ and two fermions with moduli
independent Yukawa couplings $\lambda_1$, $\lambda_2$. In this simple
model the low energy couplings at the Planck scale $M_P$ are simply
computed to be  \bea
\hat \lambda_1 \sim \left( {T_1 + T_1^+ \over T_2 + T_2^+} \right)^{3/4}
\lambda_1, \;\;\; \hat \lambda_2 \sim \left( {T_2 + T_2^+ \over T_1 +
T_1^+} \right)^{3/4} \lambda_2. \label{eq:Yuk}
\eea
Thus $\hat \lambda_1$ and $\hat \lambda_2$ are homogeneous functions of
zero degree in the moduli.

The moduli fields which describe the size and the shape of the
six-dimensional compact manifold correspond to flat directions in the
effective four-dimensional supergravity theory. If these flat
directions are exact, then the couplings in (\ref{eq:Yuk}) can be
regarded as dynamical variables to be determined by the low energy
physics (much in the spirit of the no-scale idea \cite{CFKN} used in the
dynamical determination of the gravitino mass \cite{ELNT}). It is easily
seen however that the product
\begin{equation}
\hat \lambda_1 \hat \lambda_2 \sim \lambda_1 \lambda_2 \label{eq:const}
\end{equation}
should be regarded as a constraint, because the moduli dependence has
disappeared in the right hand side of (\ref{eq:const}). Minimization of
the vacuum energy at a low energy scale with respect to the top
and bottom Yukawa couplings subject to a constraint of the type
(\ref{eq:const}) was studied in detail in \cite{BD}. It was shown there
that, qualitatively, the ratio of the two couplings behaves as
\begin{equation}
\left( {\lambda_b \over \lambda_t} \right) (\mu_0) \sim g^4 (M_P) {\mu
\over M_{SUSY}},
\end{equation}
wher $\mu_0 \sim 1 \; TeV$, $g(M_P)$ is the gauge coupling constant at
the Planck scale, $\mu$ is the usual supersymmetric mass parameter of
the MSSM and $M_{SUSY}$ is the typical mass splitting between
superpartners. For a large region of the parameter space of the MSSM,
one can thus obtain $\lambda_t / \lambda_b \sim 40-50$ and easily fit
the experimental masses with values of $\tan \beta$ of order 1.

The purpose of the present paper is twofold. First of all, we wish to
show that in effective superstrings of the orbifold type \cite{DHVW},
structures of the type (\ref{eq:X}) are naturally obtained. In this
approach, the small parameter $\varepsilon = <\phi> / M_X$ of the
$U(1)_X$ horizontal symmetry is given here by $\varepsilon = (T_1 +
T_1^+) /  (T_2 + T_2^+)$ in the case of two moduli. For $n$ moduli, we
obtain potentially $n-1$ small parameters and structures similar to the
one given by a $[U(1)_X]^{n-1}$ horizontal symmetry.

The second goal is to show that one of the cases leading to these
Froggatt-Nielsen structures\footnote{ From now on, we will call
Froggatt-Nielsen structures mass matrices for which the order of
magnitude of all entries are rational powers of a small common quantity
$\varepsilon$.} corresponds precisely to Yukawa couplings being
homogeneous functions of the moduli. We can then apply the
above-mentionned dynamical mechanism and determine by minimization the
whole structure of the fermion matrices. The hierarchy translates into
different vacuum expectation values of the moduli fields and different
modular weights of the fermions with respect to these moduli.

Section 2 presents all the cases corresponding to Froggatt-Nielsen
structures in orbifold-like effective models. In some instances, they
appear when some -- but not all -- moduli fields are fixed to their
self-dual values $T_i = 1$. An appealing situation is the case where the
theory possesses a ``diagonal'' modular symmetry; then the Yukawa
couplings are homogeneous functions of zero degree in the moduli and
Froggatt-Nielsen structures appear even if all $T_i$ are different from
1.

Section 3 analyses, in analogy with the $U(1)_X$ approach, the
relation between the mass matrices and the one-loop modular anomalies.
It is shown that if there are no string threshold corrections in the
gauge coupling constants the anomalies can be eliminated only by the
Green-Schwarz mechanism \cite{GS} which uses the Kalb-Ramond
antisymmetric tensor field present in superstring theories. In the
case relevant for the dynamical approach, the modular anomalies can be
cancelled by this mechanism only if there exists at least two moduli
with modular anomalies cancelled by the Green-Schwarz mechanism.
If the threshold corrections are present, they can account for a part
of the modular anomalies and can provide a correct gauge coupling
unification scenario, provided the modular weights of the Higgs fields
satisfy a certain relation.

Section 4 deals with the dynamical determination of the mass matrices
at low energy along the lines of ref. \cite{BD}. Two additional
constraints on the modular weights are needed for the mechanism to be
effective.

Section 5 is dedicated to a search of realistic orbifold models. It is
found that in all the cases leading to Froggatt-Nielsen
structures, no model can be constructed at Kac-Moody level one.
Possible solutions exist at level two and three, even in the simplest
case of only one small parameter. In the models with ``diagonal''
modular symmetry, we must appeal to more small parameters or,
alternatively, go to higher Kac-Moody levels. Explicit examples with two
small parameters at level three are given.

Some conclusions are presented at the end, together with open
questions that remain to be investigated.

\section {Low-energy mass matrices.} \vskip .5cm
The low energy limit of the superstring models relevant for the phenomenology
is the $N=1$
supergravity described by the K\"ahler function $K$, the superpotential $W$ and
the gauge
kinetic function $f$ \cite{CFGP} . The generic fields present in the zero-mass
string spectrum
contain an universal dilaton-like field $S$, moduli fields generically denoted
by $T_\alpha$
(which can contain the radii-type moduli $T_\alpha$ and the complex
structure moduli  $U_\beta$) and
some matter chiral fields $\phi^i$, containing the standard model particles.
The K\"ahler potential
and the superpotential read
\bea
K &=& K_0 + \sum_i \prod_{\alpha}
t^{n^{(\alpha)}_i}_{\alpha} |\phi^i |^2 + \cdots , \nn \\
K_0 &=& {\hat K}_0(T_\alpha , T^+_\alpha ) - \ln (S+S^+) ,
\label{eq:ar1}\\ W &=& {1 \over 3} \lambda_{ijk} \phi^i \phi^j \phi^k +
\cdots , \nn \eea
where the dots stand for higher-order terms in the fields $\phi^i$. In \
(\ref{eq:ar1}), $t_\alpha =
Re T_\alpha$ are the real parts of the moduli and $n_i^{(\alpha )}$ are called
the
modular weights of the fields $\phi^i$ with respect to the modulus
$T_\alpha$ . The $\lambda_{ijk}$ are the  Yukawa couplings which
may depend nonperturbatively on $S$ and $T_\alpha$. We define the
diagonal modular weight of the field $\phi^i$ as \bea
n_i = \sum_\alpha n_i^{(\alpha)} . \label{eq:ar2}
\eea
An important role in the following discussion will be played by the
target-space modular
symmetries\ $SL(2,\Z)$\ associated with the moduli fields $T_\alpha$ , acting
as
\bea
T_\alpha \rightarrow {a_\alpha T_\alpha - i b_\alpha \over i c_\alpha
T_\alpha + d_\alpha} ,\; a_\alpha d_\alpha - b_\alpha c_\alpha = 1 ,\;
a_\alpha \cdots d_\alpha \in \Z \;\; . \label{eq:ar3}
\eea
In effective string theories of the orbifold type \cite{DHVW}, the matter
fields \ $\phi^i$ transform under
(\ref{eq:ar3}) as \bea
\phi^i \rightarrow  (i c_\alpha T_\alpha + d_\alpha )^{n_i^{(\alpha)}}
\phi^i  \label{eq:ar4}
\eea
in order for the K\"ahler metric $K_i^j = \partial^2 K / \partial
\phi^i \partial \bar \phi_j$ to be invariant.

This can be viewed as a particular type of K\"ahler transformations,
which are symmetries of the supergravity theory. Denoting by $z$ the set
of the chiral fields, they read
\bea
K(z,z^+) &\rightarrow &K(z,z^+) + F(z) + F^+(z^+) , \nn\\
 &&\label{eq:ar5} \\
W(z) & \rightarrow & e^{-F(z)} \ W(z) , \nn
\eea
where $F(z)$ is an analytic function.

A typical example is a model with $n$ moduli fields and
K\"ahler potential
\bea
K_0 = - {3 \over n} \sum_{\alpha = 1}^n \ln (T_\alpha + T_\alpha^+ ) -
\ln (S+S^+) . \label{eq:ar6} \eea
Under (\ref{eq:ar3}), it transforms as
\bea
K \rightarrow K + {3 \over n} \ln | i c_\alpha T_\alpha + d_\alpha |^2
\label{eq:ar7} \eea
and the identification between eqs.(\ref{eq:ar5}) and  (\ref{eq:ar7})
gives $F_\alpha = {3 \over n} \ln (ic_\alpha T_\alpha + d_\alpha )$.
Associating a modular weight $n^{(\alpha )}_{ijk}$ with the trilinear
couplings in eq.(\ref{eq:ar1}), the transformation (\ref{eq:ar5}) of $W$
gives
\bea
n^{(\alpha )}_i + n_j^{(\alpha )} + n_k^{(\alpha)} +
n^{(\alpha )}_{ijk} = - {3 \over n} . \label{eq:ar8}
\eea
Taking the sum of all such relations for the moduli fields, we find
\bea
n_i + n_j + n_k + n_{ijk} = - 3 . \label{eq:ar9}
\eea
The factor $-3$ in the right-hand side is related to  the no-scale
structure of the K\"ahler potential (\ref{eq:ar6}) and hence to the
vanishing of the cosmological constant at tree level. More generally,
$T^\alpha K_\alpha = - 3 - V_0$, where $V_0$ is a nonzero contribution
to the cosmological constant and the right-hand side is replaced by $-3
- V_0$. Eq.(\ref{eq:ar9}) is a weaker form of eqs.(\ref{eq:ar8}),
expressing the invariance of the theory under the diagonal modular
transformations with respect to all moduli~:
\bea
\phi^i \rightarrow \prod_\alpha (ic_\alpha T_\alpha + d_\alpha
)^{n_i^{(\alpha)}} \phi^i. \label{eq:ar10'}
\eea
The difference between the
individual modular transformations and the less restrictive diagonal one
will be essential in the following.

The low energy spontaneously broken theory contains the canonically
normalized field $\hat \phi^i$ defined by $\phi^i = (K^{-1/2})^i_j \
\hat{\phi^j}$ and the Yukawas $\hat \lambda_{ijk}$ which give the
physical masses. The matching condition at the Planck scale $M_p$
relating the low energy and the original Yukawa couplings is \bea
\hat \lambda_{ijk} = e^{K_0 \over 2} (K^{-1/2})_i^{i'} (K^{-1/2})_j^{j'}
(K^{-1/2})_k^{k'}
\lambda_{i'j'k'} \ . \label{eq:ar10}
\eea
{}From eq.(\ref{eq:ar10}) we see that the $\hat \lambda_{ijk}$ are
functions of the moduli through the K\"ahler potential $K$ and
eventually the $\lambda_{i'j'k'}$.

Our goal is to analyze the general structure of the mass matrices for
the quarks and leptons as a
function of the moduli fields. They are described by the superpotential
$\hat W$ of the Minimal Supersymmetric Standard Model (MSSM) which we
take to be the minimal model obtained in the low-energy limit of the
superstring models, plus eventually some extra matter singlet under the
Standard Model gauge group. $\hat W$ contains the Yukawa interactions
\bea
\hat W \supset \hat \lambda_{ij}^U \hat Q^i \hat U^c_j \hat H_2 +
\hat \lambda^D_{ij} \hat Q^i \hat D^c_j \hat H_1 + \hat \lambda^L_{ij}
\hat L^i \hat E^c_j \hat H_1 , \label{eq:ar11}
\eea
where $H_1$ and $H_2$ are the two Higgs doublets of MSSM , $Q^i$, $L^i$ are the
$SU(2)$ quark
and lepton doublets and $U_j$, $D_j$ , $E_j$ are the right-handed
$SU(2)$ singlets.

Consider the case of two moduli $T_1$ and $T_2$. Using eqs.
(\ref{eq:ar1}), (\ref{eq:ar2}),
(\ref{eq:ar10}) and (\ref{eq:ar11}) , a $U$-quark coupling reads
\bea
\hat \lambda^U_{ij} = e^{K_0 \over 2} \ t_2^{- {n_{Q_i} + n_{U_j} + n_{H_2}
\over 2}} {\big ({t_1
\over t_2}\big ) ^{- {n_{Q_i}^{(1)} + n_{U_j}^{(1)} + n_{H_2}^{(1)}
\over 2}}} \lambda^U_{ij}\ \label{eq:ar12}
\eea
or equivalently,
\bea
\hat \lambda^U_{ij} = e^{K_0 \over 2} \ t_1^{- {n_{Q_i} + n_{U_j} + n_{H_2}
\over 2}} {\big ({t_2
\over t_1}\big ) ^{- {n_{Q_i}^{(2)} + n_{U_j}^{(2)} + n_{H_2}^{(2)} \over 2}}}
\lambda^U_{ij}
\ . \label{eq:ar13}
\eea
{\em Suppose that one of the two moduli-dependent factors in
(\ref{eq:ar12}) (or equivalently  (\ref{eq:ar13})) happens to be family
blind. Then the structure obtained for the Yukawa matrix turns out to be
very similar to the one that would be derived from an horizontal $U(1)$
symmetry  of the Froggatt-Nielsen type \cite{FN}. Modular weights play
the role of the $U(1)$ charges.} Such a situation may arise in the
following three cases of interest:

i) $t_1 = t_2 = t \not= 1$.

Then
\bea
\hat \lambda^U_{ij} =   t^{- {n_{Q_i} + n_{U_j} + n_{H_2} + 3 \over 2}}
\ \lambda^U_{ij}
\ , \label{eq:ar130}
\eea
where $n_{Q_i}$, etc are the diagonal modular weights defined in eq.
(\ref{eq:ar2}). For $t << 1$ this could produce hierarchical Yukawa couplings.
This case
is disfavoured in the case where the relation (\ref{eq:ar9}) holds with
$n_{ijk} = 0$.

ii) $t_2 = 1 , {t_1 \over t_2} = \varepsilon <<  1 $ or vice versa $t_1
\leftrightarrow t_2$.
Then using eq.(\ref{eq:ar12}) we get
\bea
\hat \lambda^U_{ij} \sim
\varepsilon^{-{ n_{Q_i}^{(1)} + n_{U_j}^{(1)} + n_{H_2}^{(1)} \over 2}}
\lambda^U _{ij} , \label{eq:ar14}
\eea
where we dropped the universal $e^{K_0 \over 2}$ factor, irrelevant
here. Hierarchical structures are obtained if the dynamics imposes
$\varepsilon = {t_1 \over t_2}$ small (typically of the order of the
Cabibbo angle to some power). Remark that the relevant modular weights
correspond to  the modulus whose ground state falls away from the
self-dual points, $t_i \not= 1$ (for an example of such
a situation, see Ref.\cite{MacRoss}).

iii) one has the condition
\bea
n_{Q_i} + n_{U_j} + n_{H_2} = \ \mbox{independent of }\  i\  \mbox{and}
\ j . \label{eq:ar15}
\eea
and ${t_1 \over t_2} = \varepsilon <<  1 $ or vice-versa.
This obviously implies that $n_{Q_i} = n_Q$ and $n_{U_i} = n_U$
for any $i = 1, 2, 3$.

For example, in the case when the diagonal modular symmetry holds, the
constant (\ref{eq:ar15}) is equal to $-3$ and the Yukawa couplings can be
written as
\bea
\hat \lambda_{ij}^U = {1 \over S + \bar S}
\left( {t_1 \over t_2} \right)^{-{ n_{Q_i}^{(1)} + n_{U_j}^{(1)} +
n_{H_2}^{(1)} + {3 \over 2} \over 2}}  \lambda_{ij}^U \; = \;
{1 \over S + \bar S} \left( {t_2 \over t_1} \right)^{-{
n_{Q_i}^{(2)} + n_{U_j}^{(2)} + n_{H_2}^{(2)} + {3 \over 2} \over 2}}
\lambda_{ij}^U \label{eq:ar15a}
\eea
or
\bea
\hat \lambda_{ij}^U = {1 \over S + \bar S} \left( {t_1 \over t_2}
\right)^{-{ n_{Q_i}^{(12)} + n_{U_j}^{(12)} + n_{H_2}^{(12)} \over
4}}  \lambda_{ij}^U \label{eq:ar15b}
\eea
where $n_{Q_i}^{(12)} = n_{Q_i}^{(1)} - n_{Q_i}^{(2)}$, etc. The last
form (\ref{eq:ar15b}) is particularly useful in that it relates the
Froggatt-Nielsen-like structures with the asymmetry between the modular
weights corresponding to the two moduli fields.
\footnote{ Such formulas show that the magnitude of the modular
weights must appear in a hierarchy which is opposite for the two
moduli. In fact, the construction of models with two moduli  along these
lines leads to a difficulty compared to the case i). This is because in
order to get the required hierarchy, for example
\bea
\hat \lambda^U_{ij} / \hat \lambda^U_{33} \ \sim \  \varepsilon^{-{
n^{(1)}_{Q_i} - n^{(1)}_{Q_3} + n_{U_j}^{(1)} - n_{U_3}^{(1)} \over 2}} \
\lambda^U_{ij}  /\lambda^U_{33} , \label{eq:ar17} \eea
we need\  $| n_{Q_i}^{(1)} - n_{Q_3}^{(1)} + n_{U_j}^{(1)} - n_{U_3}^{(1)} |
>> 1$. But the condition
(\ref{eq:ar15}) implies\  $n_{Q_i}^{(1)} - n_{Q_3}^{(1)} + n_{U_j}^{(1)}
- n_{U_3}^{(1)} = - \big (n_{Q_i}^{(2)} - n_{Q_3}^{(2)} + n_{U_j}^{(2)} -
n_{U_3}^{(2)} \big ) $\ so a large asymmetry of the modular weights for
the moduli\  $T_1$\  should be compensated by a large asymmetry, of
opposite sign, coming from the modular weights of the moduli \ $T_2$.
This is difficult to satisfy and asks generally for more that two moduli
or (and) higher Kac-Moody levels, as we will see later in explicit
constructions.
But the construction above is easily generalized to the case of 3 or
more moduli.}

%

One way to get the condition (\ref{eq:ar15}) is to search for models
where the $\hat \lambda_{ijk}$ are \underline{homogeneous functions} of
the moduli $T_\alpha$, i.e. $\sum_\alpha T_\alpha \partial \hat
\lambda_{ijk} / \partial T_\alpha = 0$. In this case, using the
relation $\sum_\alpha t_\alpha \partial  K^j_i / \partial t_\alpha
= n_i K^j_i$ and the matching condition (\ref{eq:ar10}), we arrive at an
equation for the original  couplings $\lambda_{ijk}$
\bea
\big ( {1 \over2} T_\alpha K^\alpha - {n_i + n_j + n_k \over 2} + T_\alpha
{\partial \over \partial
T_\alpha} \big ) \lambda_{ijk} = 0 . \label{eq:ar16}
\eea
If $T_\alpha K^\alpha = - 3$ and the $\lambda_{ijk}$ are pure
numbers we recover  eq.(\ref{eq:ar9}). In such a case ($n_{ijk} =
0$), the relation (\ref{eq:ar16}) can be derived from assuming the
diagonal modular symmetry discussed above in (\ref{eq:ar10'}). This
approach was used in \cite{BD} in a dynamical approach to the fermion
mass problem proposed in \cite{N}, \cite{Z}  and studied in \cite{KPZ}
and \cite{BDu}. We will return to it in section $4$.

The experimental data on the fermion and the mixing angles can be summarized as
follows.
Defining $\lambda = sin \theta_c \sim 0.22$ where $\theta_c$ is the Cabibbo
angle, the mass
ratios and the Kobayashi-Maskawa matrix elements at a high scale $M_X
\sim M_P$ have the values \bea
&{m_u \over m_t} \sim \lambda^7 \; {\rm to} \; \lambda^8 \ ,
\  {m_c \over m_t} \sim \lambda^4 \ ,
\  {m_d \over m_b} \sim \lambda^4\  ,
\  {m_s \over m_b} \sim \lambda^2\ ,\nn \\
&{m_e \over m_\tau } \sim \lambda^4 \ ,
\ {m_\mu \over m_\tau } \sim \lambda^2 \ ,
\ | V_{us} | \sim \lambda \ ,
\ | V_{cb} | \sim \lambda^2 \ ,
\ | V_{ub} | \sim \lambda^3 \; {\rm to} \;  \lambda^4
. \label{eq:ar19}
\eea
Taking as a small parameter $\varepsilon =
\lambda^2 \sim {1 \over 20}$, these values are perfectly
acommodated by the following modular weight assignement
\bea
n_{Q_3}^{(1)} - n_{Q_1}^{(1)} = 3 \ , \ n_{Q_3}^{(1)} - n_{Q_2}^{(1)} =
2 \nn \\ n_s^{(1)} - n_b^{(1)} = 0 \ , \ n_b^{(1)} - n_d^{(1)} = 1
\label{eq:ar20} \\ n_t^{(1)} - n_c^{(1)} = 2 \ , \ n_t^{(1)} - n_u^{(1)}
= 4 \nn
\eea
corresponding to the mass matrices
\bea
\hat \lambda_U = {\lambda^{-x}} \left ( \begin{array}{cc}
\lambda^7 \ \lambda^5 \   \lambda^3    \\
\lambda^6 \   \lambda^4  \  \lambda^2    \\
\lambda^4 \  \lambda^2 \ \ 1
\end{array}  \right ) , \;\;
\hat \lambda_D = {\lambda^y} \left ( \begin{array}{cc}
\lambda^4 \ \lambda^3 \ \lambda^3   \\
\lambda^3 \  \lambda^2 \ \lambda^2   \\
\lambda \ \ 1 \ \ 1
\end{array}  \right ) . \label{eq:ar21}
\eea
In (\ref{eq:ar21}) $x \ge 0$ (the top coupling should be at least of
order one at a high scale) and $y \ge 0$ (the bottom coupling should
correspondingly be smaller or equal to one). Remark that the negative
power of $\lambda$ in $\hat \lambda_U$ is impossible to obtain in a
horizontal symmetry approach because of the analyticity of the
superpotential. In the moduli case it comes naturally and it plays an
important role in a dynamical approach to the fermion masses, as will
be shown in section $4$.

A detailed analysis in the case of an horizontal $U(1)$ symmetry  shows
that other assignments which fit approximately the values
(\ref{eq:ar19}) are possible \cite{DPS} , but the one considered here is
the best suited for our purposes. This is because the others ask for
larger modular weight differences, which are difficult to get in
realistic orbifold models. This topic will be discussed in greater
detail in section $5$. The expressions (\ref{eq:ar21}) will be the
starting point in the construction of abelian orbifold models with
realistic mass matrices.

An interesting aspect of the case iii) discussed above should be
stressed which concerns the sfermion masses $(M^2_0)_{i \bar \jmath}$.
Non-diagonal sfermion mass matrices give potentially
dangerous contributions to flavor changing neutral current processes
like $b\rightarrow s \gamma$ or $\mu \rightarrow eJ\gamma$ \cite{GM}. The
general expression in supergravity is  \bea (M^2_0)_{i \bar \jmath} =
(M_{1/2} M^+_{1/2})_{i \bar \jmath} + (G_{i \bar \jmath} - G_\alpha
R^\alpha_{{\bar \jmath}i \beta } G^\beta ) m_{3/2}^2 , \label{eq:ar22}
\eea
where $G = K + l n | W |^2$ and $ G_{i \bar \jmath} = {\partial^2 G
\over \partial \phi^i \partial \phi^+_{\bar \jmath}} $
is the metric on the K\"ahler space.
The indices $\alpha , \beta $ correspond to moduli fields which contribute to
supersymmetry
breaking $< G_\alpha > \not= 0$ and we assume $< G_\alpha G^\alpha > = 3. $
(This corresponds to
the moduli limit in the language of refs \cite{BIM}, \cite{FKZ} where
the universal dilaton does not contribute to supersymmetry breaking. The
general case is irrelevant for the present analysis). $R^\alpha_{{\bar
\jmath}i\beta}$ is the Riemann tensor of the K\"ahler space,
$(M_{1/2})_{ik}$ the fermion mass matrix and $m_{3/2}$ the gravitino
mass. Using the intermediate formula
\bea
G_\alpha R^\alpha_{{\bar \jmath}i\beta } G^\beta =
t^\alpha t^\beta {\partial^2 G_{i{\bar \jmath}} \over \partial t^\alpha
\partial t^\beta } - {\partial G_{i{\bar k}} \over \partial t^\alpha }
(G^{-1})^{{\bar k }l } {\partial G_{l {\bar \jmath}} \over \partial
t^\beta } = -n_i G_{i {\bar \jmath}} \label{eq:ar23}
\eea
and doing a
trivial rescaling in order to define the low-energy parameters, we find
\bea
(M_0^2)^{U ,D}_{LL} = m_{3/2} ^2 (1 + n_Q) \1 + (M_{1/2} M^+_{1/2})^{U,D}
,\nn \\
(M_0 ^2 )^U_{RR} = m_{3/2} ^2 (1 + n_U) \1 + (M_{1/2} M^+ _{1/2})^U ,
\nn \\
(M_0 ^2 )^D_{RR} = m_{3/2} ^2 (1 + n_D) \1 + (M_{1/2} M^+ _{1/2})^D ,
\label{eq:ar24}
\eea
where $(M^2_0)^{U,D}_{LL}$ is the left-left squark
squared mass  respectively for the $U$ and the $D$ quark, etc. In
(\ref{eq:ar24}) $n_Q , n_U$ and $n_D$ are the diagonal modular weights
which in the case iii) discussed above are the same for the three
generations. The important aspect of (\ref{eq:ar24}) is that the squark
soft-breaking mass matrix is proportional to the identity. Going to the
basis where the quark mass matrices are diagonal, we see that the soft
squark masses are still proportional to the unit matrix. Consequently,
there are no flavor changing neutral currents  induced at the
supergravity level. Even if we know by now \cite{BIM} that SUGRA -
induced flavor changing neutral currents are not as severe as thought
several years ago, it is still worth emphasizing the virtue of the case
iii), corresponding to considering Yukawa as being homogeneous functions
of the moduli.  \vskip 1cm  \section {Modular anomalies and moduli mass
textures.}  \vskip .5cm
In the context of horizontal abelian symmetries used to explain fermion
mass hierarchies, an interesting connection has been established
\cite{IR,BR} between anomalies associated with such symmetries and mass
hierarchies as given in (\ref{eq:ar19}).

Gauge anomalies in usual field theories must be absent in
order to define a consistent quantum theory. In effective supergravity
theories the gravitational anomalies must cancel too in order to
preserve the reparametrization invariance. This requirement  imposes
non-trivial constraints on the particle spectrum in chiral theories.
Applied to the ten-dimensional superstrings, this led to the famous
Green-Schwarz anomaly cancellation mechanism involving the Kalb-Ramond
antisymmetric tensor field \cite{GS}. This mechanism has a counterpart
in 4 dimensions which allows to fix the value of $\sin ^2 \theta_W$ at
the string scale without advocating a grand unified symmetry \cite{IB}.
It was shown in \cite{BR} and generalized in \cite{DPS,Nir} that, using
this mechanism, it is possible to infer from the observed hierarchies
(\ref{eq:ar19}) in the mass matrices the standard value of $3/8$ for
$\sin ^2 \theta_W$.

Effective string models also have another type of anomalies,
named $\sigma$-model anomalies \cite{MN}. They appear, as in the gauge
case, in triangle diagrams with two gauge bosons and one modulus
and have two different origins: one is the nontrivial metric of the
matter fields; the other  non-invariance can
be analyzed as a violation of the K\"ahler invariance of the SUGRA theory
i.e. under transformations of the type (\ref{eq:ar3}). It is known that
modular transformations are symmetries to all orders in the string
perturbation theory. Moreover, the string massive spectrum is separately
invariant, as can be checked by interchanging the Kaluza-Klein and the
winding states. Thus it is within  the zero mass spectrum of the string,
which defines the effective SUGRA theory, that the corresponding
triangle anomalies must cancel at the field theory level. We will show
in this section that the cancellation of these anomalies plays a role
very similar to the one of mixed gauge anomalies in the
abelian horizontal $U(1)_X$ symmetry approach.
\vskip .5cm
Consider the
non-linear $\sigma$-model corresponding to the moduli fields,
generically denoted by $T_\beta$, which refers as above both to the
K\"ahler (radii-type) and complex structure moduli. The gauge group is
$G = \prod_a G_a$ and there are matter fields in different
representations $R_a$ of $G_a$. The anomalous triangle diagrams give a
non-local contribution to the one-loop effective lagrangian \cite{DFKZ}
which reads  
\bea
{\cal L}_{nl } &=& {1 \over 8} {1 \over 16 \pi^2} \sum_a
\int d^4 \theta  (W^\alpha W_\alpha)_a {{\cal D}^2 \over \Box } \ \left(
  C(G_a )   K(T_\beta , T^+_\beta ) \right. \nn \\
& & \left. +  \sum_{R_a} T(R_a) \left[ 2\ln \det K_{i\bar \jmath}^{R_a}
(T_\beta , T^+_\beta ) -  K(T_\beta , T^+_\beta ) \right] \right)
 + h.c. \label{eq:ar25}
\eea
In eq.(\ref{eq:ar25}) where superfield notations are used, $W^\alpha$ is
the Yang-Mills field strength superfield ; $K_{i\bar \jmath}^{R_a}$ is
the K\"ahler metric for the matter fields in the representation $R_a$ of
group $G_a$.

The cases of interest to be analyzed in this paper are orbifold
compactifications. Consider the diagonal K\"ahler moduli for which
$\hat K_0 (T_\beta, T^+_\beta) = - \ln (T_\beta + T^+_\beta)$ (and
possibly the complex structure moduli). Then the above expressions
reduce to
\beq
{\cal L}_{nl } = {1 \over 8} {1 \over 16 \pi^2} \sum_a
\int d^4 \theta  (W^\alpha W_\alpha)_a {{\cal D}^2 \over \Box }
\sum_\beta b_a'^{(\beta)} \ln (T_\beta + T^+_\beta )
 + h.c. \label{eq:ar26}
\eeq
where the anomaly coefficients $b_a'^{(\beta)}$ are given by the
expressions
\bea
b_a'^{(\beta)}= - C(G_a) + \sum_{R_a} T(R_a) (1 + 2 n^{(\beta)}_{R_a} )
.  \label{eq:ar27}
\eea

One finds that the change of ${\cal L}_{nl }$ under the modular
transformations (\ref{eq:ar3}) is given by the local expression
\bea
\delta {\cal L}_{nl } = {1 \over 2} {1 \over 16 \pi^2 }  \sum_a
\int d^2 \theta (W^\alpha W_\alpha)_a \sum_\beta  b_a'^{(\beta)}
\ln (i c_\beta T_\beta + d_\beta ) + h.c. \label{eq:ar28}
\eea

There are two ways of compensating this anomaly. The first, which  is
particularly interesting
in our case is reminiscent of the Green-Schwarz mechanism. It
uses  the form of the tree level gauge kinetic term
\beq
{\cal L}_{tree} = \sum_a \int d^2 \theta {1 \over 4} k_a S
(W^\alpha W_\alpha)_a + h.c. \label{eq:ar281}
\eeq
and requires the
non-invariance of the dilaton field under the modular transformations
\bea S \rightarrow S - {1 \over 8\pi^2} \sum_\beta \delta^{(\beta)}_{GS}
\ln (i c_\beta T_\beta + d_\beta ) . \label{eq:ar29}
\eea
The factor $\delta_{GS}^{(\beta)}$ is the gauge group independent Green-Schwarz
coefficient
and induces a mixing between the dilaton $S$ and the moduli fields
$T_\beta$. This mechanism can completely cancel the anomalies  only if
the anomaly coefficients $b_a'^{(\beta)}$ satisfy the equalities
\bea
\delta_{GS}^{(\beta)} = {b_a '^{(\beta )} \over k_a} = {b_b'^{(\beta)} \over
k_b} = \cdots \label{eq:ar30}
\eea
for all the group factors of the gauge group $G = \prod_a G_a$.

A second mechanism for the cancellation of the term (\ref{eq:ar28}) uses the
one-loop threshold
corrections to the gauge coupling constants, which can be different for
different gauge group
factors. They appear for the moduli fields associated with complex planes
left unrotated by some twist vectors and are due to contributions from the
massive Kaluza-Klein and winding states. If the modular symmetry group
is  $(SL(2,\Z))^3$, the one-loop running gauge coupling constants
at a scale $\mu$ reads
\bea
{1 \over g^2_a (\mu)} = {k_a \over g^2_s} + {b_a \over 16 \pi^2} ln {M^2_s
\over \mu^2} - {1 \over 16 \pi^2} \sum_{\alpha =1}^3 (b'^{(\alpha)}_a - k_a
{\delta
^{(\alpha)}_{GS}}) \ln \left [ (T_{\alpha} + T^+_{\alpha})
|\eta (T_{\alpha})|^4 \right ]  \label{eq:ar131}
\eea
In (\ref{eq:ar131}) $g_s$ is the string coupling constant, $M_s$ is the string
scale and $b_a$ are the RG $\beta$-function coefficients ($a = 1,2,3$)  for
$U(1)_Y$, $SU(2)_L$ and $SU(3)$ respectively. The
Dedekind function is defined by \ $\eta (T) = exp { (- \pi T / 12)
} \prod_{n=1}^{ \infty} [1 - exp {(2 \pi n T)}]$, which transforms under
(\ref{eq:ar3}) as  $\eta (T_{\alpha}) \rightarrow \eta (T_{\alpha}) (i
c_{\alpha} T_{\alpha} +  d_{\alpha})^{1 \over 2}$.
The unification scale $M_U$ is determined by the condition
\beq
k_1 g_1^2 (M_U) = k_2 g_2^2 (M_U) = k_3 g_3^2 (M_U), \label{eq:ar1311}
\eeq
and computed to be
\bea
M_U = M_s \prod_{\alpha =1}^3  \left [ (T_{\alpha} + T^+_{\alpha})
| \eta (T_{\alpha})|^4 \right ]^{b'^{(\alpha)}_b k_a - b'^{(\alpha)}_a k_b
\over 2
(b_a k_b - b_b k_a)} \ . \label{eq:ar132}
\eea
with $a \not= b \in \{ 1,2,3\}$.
In the simplest approximation of neglecting all the threshold corrections, a
one-loop RG analysis for $g_3$ and $sin^2 \theta_W$ gives a good agreement with
the experimental data if $M_U \simeq M_s / 50$ \cite{CEFNZ}.

Consider now a minimal orbifold model with the particle content of
the \linebreak MSSM (respectively $Q_i,U_i,D_i,L_i$, $i$ being a family
index, and the two Higgs supermultiplets $H_1$ and $H_2$), plus possibly
extra Standard Model singlet fields. The mixed K\"ahler $SU(3) \times
SU(2) \times U(1)_Y$ triangle anomalies are described by the
coefficients \cite{IL}
\bea
&b_1 '^{(\beta )} &= 11 + \sum^3_{i=1} ({1 \over 3} n^{(\beta )}_{Q_i } + {8
\over 3} n^{(\beta
)}_{U_i} + {2 \over 3} n^{(\beta )}_{D_i} + n^{(\beta )}_{L_i} + 2n^{(\beta
)}_{E_i} ) + n^{(\beta
)}_{H_1} +  n^{(\beta )}_{H_2} , \nn   \\
&b_2 '^{(\beta )} &= 5 + \sum^3_{i=1} (3n^{(\beta )}_{Q_i } + n^{(\beta
)}_{L_i}) + n^{(\beta
)}_{H_1} + n^{(\beta )}_{H_2} , \label{eq:ar31} \\
&b_3 '^{(\beta )} &= 3 + \sum^3_{i=1}
(2n^{(\beta )}_{Q_i } + n^{(\beta )}_{U_i} + n^{(\beta )}_{D_i} ). \nn
\eea
Apart from the modular weight independent piece, these coefficients
are identical to the ones encountered for the
mixed $U(1)_X - G_a$ gauge group anomalies in the abelian horizontal
$U(1)_X$ gauge symmetry approach. Again, the role of the $U(1)_X$ charges
is played here by the modular weights of the different fields.
Consequently we will closely follow  the analysis performed in  \cite{IR}
, \cite{BR} and \cite{DPS}.

We first place ourselves in the case ii) of the preceding
section, where only one modular weight is relevant for the mass
matrices. In what follows, the modular weights and the anomaly
coefficients refer to only one of the moduli. As noted in Ref.
\cite{BR} in the case of an horizontal abelian symmetry, an interesting
consequence of eqs.(\ref{eq:ar14}) and (\ref{eq:ar31}) is
\bea
{ Det
\hl_L \over Det \hl_D } \sim \varepsilon^{2 + {n_{H_1} + n_{H_2} \over
2} - {1 \over 4} (b'_1 + b'_2 - {8 \over 3} b'_3 ) } \ .\label{eq:ar32}
\eea
Also, as in the analysis of \cite{DPS,Nir}, we have
\bea
 ({\rm Det} \hl_U ) ({\rm Det} \hl_L)^3 ({\rm Det} \hl_D)^{-2} \sim
\varepsilon^{-{3 \over 4} (b'_1 + b'_2 - 2b'_3 - 10)} \ .
\label{eq:ar33}
\eea
This equation has the advantage not to contain the
unknown variable $n_{H_1} + n_{H_2} $, which allows to draw general
conclusions. It is clear from (\ref{eq:ar33}) that for
realistic mass values and phenomenologically acceptable ratio
$\lambda_b / \lambda_t$ we cannot put the anomalies to zero $b'_1
= b'_2 = b'_3 = 0$.

In the case where there are no threshold corrections associated with the
modulus giving the mass structures, the only solution is the use of the
Green-Schwarz mechanism  \cite{IR,BR}
\bea
b'_2 = b'_3 = {3 \over 5} b'_1 = b' \label{eq:ar34}
\eea
corresponding to Kac-Moody level ratios
\beq
k_1 : k_2 : k_3 = {5 \over 3} : 1 : 1. \label{eq:ar341}
\eeq
As is well
known, this leads to the successful prediction of the Weinberg angle at
a high scale $M_U$, $ sin^2 \theta_W = {3 \over 8}$ \cite{IB}. Using the
experimental input (\ref{eq:ar19}) which can be expressed as $Det \hl_L
\sim Det \hl_D \sim \lambda^{6+3y}$ , $Det \hl_U \sim \lambda^{12-3x}$ ,
and taking $\epsilon \sim \lambda^2$, eqs.(\ref{eq:ar32}) ,
(\ref{eq:ar33}) and (\ref{eq:ar34}) fix the variables
\bea
&n_{H_1} + n_{H_2} = -4 , \nn \\ &b' = - 3 + 3(x-y) , \label{eq:ar35}
\eea
where $x$ and $y$ are defined in (\ref{eq:ar21}) .

If there are threshold corrections in the modulus
field giving the Froggatt-Nielsen structures, then  eq.
(\ref{eq:ar32}) can be rewritten in the form (assuming the Kac-Moody
levels (\ref{eq:ar341}))
\bea
{ {\rm Det} \hl_L \over {\rm Det} \hl_D }
{\sim \varepsilon^{2 + {n_{H_1} + n_{H_2} \over 2}}}  (M_U /
M_s)^{{1 \over 2} (b_1 + b_2 - {8 \over 3} b_3 ) \ ln \varepsilon
\left [ ln (T_{\alpha} + T^+_{\alpha})| \eta (T_{\alpha})|^4 \right
]^{-1}} \ . \label{eq:ar133}
\eea
The ratio of the two logarithms in the
right-hand side is easy to compute for $t_{\alpha} \sim
\varepsilon$. In this case, $(T + T^+) | \eta (T) |^4 \sim {2 \over
\varepsilon} e^{-\pi \over 3 \varepsilon}$. Consequently the ratio of
the two logarithms is approximately $-3 \varepsilon ln \varepsilon
/\pi \sim  1 / 4$.  Because of modular invariance a
similar value is obtained for $t_\alpha \sim \varepsilon^{-1}$. For the
phenomenologically relevant case $M_U \sim {M_s}/50$, the successful
relation ${\rm Det} \hl_L = {\rm Det} \hl_D$ asks for
\bea
n_{H_1} + n_{H_2} \simeq -10 \ . \label{eq:ar134}
\eea
This is an interesting possibility which will be further investigated in
section $5$. Using the possible modular weights of the Higgs fields
at Kac-Moody levels $2$ and $3$ from the tables $1$ and $2$ and the known
renormalization group  coefficients $b_a$, we can find the allowed values
of the unification scale $M_U$. As a result,  as we will discuss in
section $5$, it is found that the relation (\ref{eq:ar134}) can
be satisfied. Thus, the threshold corrections contribute in order to give
a good unification scheme as well as realistic mass textures.

Eqs. (\ref{eq:ar134}) and (\ref{eq:ar35}) must be taken into account in
the construction of explicit models. Generically it is clear, by a
simple inspection of eqs.(\ref{eq:ar31}) that we need many fields in the
twisted sectors of the orbifolds in order to satisfy eqs.
(\ref{eq:ar134}) and (\ref{eq:ar35}). Twisted fields are in any case
necessary in order to get the assignement (\ref{eq:ar20}) leading to the
desired hierarchical structure.

We now turn to the case  (iii) of the preceding section. Starting from
the relation (\ref{eq:ar15b}), we
obtain, using the same technique as before:
\bea
({\rm Det} \hl_U ) ({\rm Det} \hl_L)^3 ({\rm Det} \hl_D)^{-2} &\sim &
\varepsilon^{-{3 \over 8} (b_1'^{(12)} + b_2'^{(12)} - 2b_3'^{(12)})},
\nn \\
{ Det \hl_L \over Det \hl_D } &\sim & \varepsilon^{-{1 \over 8}
[b_1'^{(12)} + b_2'^{(12)}  - {8 \over 3} b_3'^{(12)} - 2
(n_{H_1}^{(12)} + n_{H_2}^{(12)} )] } \ .\label{eq:n47}
\eea
The first of eqs. (\ref{eq:n47}) is very useful to discuss anomaly
cancellation conditions. Taking as an example $\varepsilon \sim
\lambda^m$, it requires
\beq
b_1'^{(1)} + b_2'^{(1)} - 2 b_3'^{(1)} = b_1'^{(2)} + b_2'^{(2)} - 2
b_3'^{(2)} - {48 \over m}. \label{eq:n48}
\eeq
As shown in \cite{BR} for the case of an horizontal symmetry, the second
of eqs. (\ref{eq:n47}) has the following interesting solution, which
automatically gives the value $3/8$ for $\sin^2 \theta_W$ at
unification:
\bea
 n_{H_1}^{(1)} + n_{H_2}^{(1)} &=& n_{H_1}^{(2)} + n_{H_2}^{(2)}, \nn \\
b_1'^{(1)} + b_2'^{(1)} - {8 \over 3} b_3'^{(1)} &=& b_1'^{(2)} +
b_2'^{(2)} - {8 \over 3} b_3'^{(2)} . \label{eq:n49}
\eea

Moreover, using the conditions (\ref{eq:ar9}) in the case $n_{ijk} = 0$
and the expressions (\ref{eq:ar31}), we obtain
\bea
b'_1 + b'_2 - 2 b'_3 &=& 8, \nn  \\
b'_1 + b'_2 - {8 \over 3} b'_3 &=& 2 (8 + n_{H_1} + n_{H_2}),
\label{eq:n50}
\eea
where $b'_1 = b_1'^{(1)} + b_1'^{(2)}$, etc. Eqs. (\ref{eq:n48}) and
(\ref{eq:n50}) clearly express the fact that the theory has one-loop
modular anomalies.

An analysis of all the possibilities for the anomalies related to the
two moduli leads to the conclusion that, without threshold corrections,
the mixed case with zero anomalies for one modulus and Green-Schwarz
mechanism for the other modulus is physically uninteresting (it
requires $\varepsilon \sim \lambda^{\pm 6}$). In the case of anomalies
cancelled by the Green-Schwarz mechanism for both moduli, we obtain
$n_{H_1}^{(1)} + n_{H_2}^{(1)} = n_{H_1}^{(2)} + n_{H_2}^{(2)}=-4$
and $b'^{(i)} = 6 ( 1 \pm 6/m)$ for $i=2, 1$. The only other allowed
case is when threshold corrections are present for both moduli. In this
case, we obtain $n_{H_1} + n_{H_2} + 8 = {60 \lambda^m \over \pi} \ln
(M_S^2 / M_U^2)$. A realistic value for $M_U$ requires $m \le 2$ and is
obtained for example for $m=2$, $n_{H_1} + n_{H_2}= -14$. Let us note
that $\sin^2 \theta_W$ can still be found equal to $3/8$ at unification
scale, irrespective of the choices made in order to obtain the desired
value for $M_U$.

\vskip 1cm

\section {Dynamical determination of couplings.}
\vskip .5cm
The duality symmetries imply the existence of flat directions in the
corresponding moduli fields. If they are respected to all orders in
the supergravity interactions, then the only way to lift them is
by breaking supersymmetry. Given the scale expected for this breaking,
one may expect the low energy sector to play an important role in the
determination of the moduli ground state. Under these conditions, the
low energy minimization with respect to the moduli fields is
presumably equivalent to the minimization with respect to the Yukawa
couplings, through their non-trivial dependence on the moduli. This was
the attitude taken in Refs. \cite{KPZ,BDu,BD} to dynamically determine
the top/bottom Yukawa couplings. A very important point in this program
is the existence of constraints between Yukawas , of a type which
is typical of the approach based on moduli dynamics. It was shown in
Ref. \cite{BD} that this can be enforced if the Yukawa couplings are
homogeneous functions of the moduli. In what follows, we will
therefore place ourselves in the case iii) of section $2$ and analyze
how the two approaches can be merged, leading to a dynamical
determination of the fermion mass hierarchies and mixing angles.

We start by reviewing the results of Ref. \cite{BD}. To
compute the vacuum energy at the low-energy scale $\mu_0 \sim M_{susy}$
we proceed in the usual way. Using boundary values compatible with the
constraints at the Planck scale $M_P$ (identified here with the
unification scale), we evolve the running parameters down to the scale
$\mu_0$ using the RG equations and adopt the effective potential
approach \cite{CW}. The one-loop effective potential has two pieces  \bea
V_1 (\mu_0) = V_0 (\mu_0) + \Delta V_1 (\mu_0) \ , \label{eq:ar101}
\eea
where $V_0 (\mu_0)$ is the renormalization group improved tree-level
potential and $\Delta V_1 (\mu_0)$ summarizes the quantum corrections
given by the formula
\bea
\Delta V_1 (\mu_0) = {(1 / 64 \pi^2)} \ Str \cm^4 \ (\ln {\cm^2 \over
\mu_0^2} - {3 \over 2}) \ . \label{eq:ar102}
\eea
In (\ref{eq:ar102}) $\cm$ is the field-dependent mass matrix,
$Str \cm^n = \sum_J (-1)^{2J} (2J+1) \ Tr M^n_J$ is the ponderated trace
of the mass matrix for particles of spin $J$  and all the
parameters are computed at the scale $\mu_0$. The vacuum state is
determined by the equation $\partial V_1 / \partial \phi_i = 0$, where
$\phi_i$ denotes collectively all the fields of the theory. The vacuum
energy is simply the value of the effective potential computed at the
minimum.\footnote{ In a first approximation, if the moduli masses are
larger than the average superpartner mass ${\tilde m}$, the factor
$\ln \cm^2 / \mu_0^2$ in (\ref{eq:ar102}) can be replaced by $\ln
{\tilde m}^2 / \mu_0^2 < 0$. \cite{BDu,BD}}

As expected there is no Yukawa coupling dependence at the tree level.
At the one-loop level it appears through
\bea
 \bal { 1 \over 3} \ Str \cm^4 =  A_U Tr \lambda^2_{U} +
A_D Tr (\lambda^2_{D} + {1 \over 3} \lambda^2_{L}) + 8 \mu Tr (\lambda_U
{\cal A_U} + \lambda_D {\cal A_D} + {1 \over 3} \lambda_L {\cal A_L}) v_1
v_2 \ , \label{eq:ar54} \eal  \eea
where $v_1$ and $v_2$ are the vacuum expectation values of the two
Higgs doublets. In (\ref{eq:ar54}) ${\cal A_U}$, ${\cal A_D}$ and ${\cal
A_L}$ are trilinear soft breaking terms and the trace is in the family
space. $A_U$ and $A_D$ are given by the expressions \bea
\bal A_U = 2 \ [ 2\mu^2 / tg^2 \beta +
4 M^2 - M^2_Z + (g^2_1 + g^2_2) v^2_1 ] \ v^2_2 \ , \\ \\ A_D = 2 \ [ 2\mu^2 \
tg^2 \beta + 4 M^2 - M^2_Z + (g^2_1 + g^2_2) v^2_2 ] \ v^2_1 \ ,
\label{eq:ar55} \eal
\eea
where  $g_1,g_2$ are the $U(1)$, $SU(2)$ gauge couplings, $M$ is a
universal squark soft mass and $M^2_Z =  {1 \over 2} (g^2_1 + g^2_2 )
(v^2_1 + v^2_2 ) $ is the $Z$  mass. In order to show that $A_U,A_D>0$,
one may use the phenomenological inequality
\bea
 (Str \cm^2)_{\mbox{quarks + squarks}} = 4M^2 > M^2_Z \ .
\label{eq:ar56}  \eea
The vacuum energy (\ref{eq:ar101}) has roughly
the Nambu form \cite{N} with
an additional linear term which does not change the shape
of the vacuum energy as a function of the Yukawas, but which plays an
essential role in the minimization process.

The positivity of $A_U$, $A_D$ is a consequence of supersymmetry
in the sense that it is due to the Yukawa dependent bosonic contributions
in (\ref{eq:ar54}). In the non-supersymmetric Standard Model the sign
is negative and the present considerations do not apply.
Using eq.(\ref{eq:ar101}) and eq.(\ref{eq:ar102}), we obtain the vacuum
energy as a function of the matrices $\lambda_U$ and $\lambda_D$, which is a
paraboloid unbounded from below. If the minimization is freely performed,
then they are driven to the maximally allowed values and no hierarchy
is generated.

Consider now the mass matrices (\ref{eq:ar21}) with
$\lambda \sim (t_1 / t_2)^{1 \over 2}$ a dynamical parameter to be
determined by the minimization.
We discussed in Ref.\cite{BD} two types of constraints: (a) the
proportionality constraint where one of the couplings is proportional
to another (to some positive power) $\lambda_1 = {\rm cst}\cdot
\lambda_2^n,\; n>0$, (b) a {\em multiplicative constraint} where the
product of two couplings (or positive powers of them) is fixed to be a
moduli independent constant: $\lambda_1 \lambda_2^n = {\rm cst}, \;
n>0$. Only the second constraint leads to dynamical hierarchy of
couplings. Fortunately for $x,y > 0$ in (\ref{eq:ar21}) we get the second
type of constraints, for example $(\hl^{33}_U)^y  (\hl^{33}_D)^x
= cst$. In this case if $\hl^{33}_U$ for example is big, the
constraint (valid at $M_P$) forces $\hl^{33}_D$ to be small and we
naturally obtain small numbers.

For the case of two moduli, the conditions to have $x > 0$, $y > 0$ read
\bea
n_{Q_3} + n_{U_3} + n_{H_2} > -3/2 \ , n_{Q_3} + n_{D_3} + n_{H_1} <
-3/2 \label{eq:ar57} \eea
and they should be fulfilled in order to obtain multiplicative-type
constraints.
An interesting case (treated in detail in \cite{BD}, where we keeped only
$\lambda^{33}_U$ and $\hl^{33}_D$ in the computations) is $x = y$.
The relevant constraints are then symmetric in the up and down quarks.

The low energy effective potential is to be minimized with respect to
$\lambda$.
For this the RG equations are used in order to translate the structures
(\ref{eq:ar21})
from $M_P$ to $\mu$. The analysis is essentially the same as in \cite{BD}, the
whole structure of the mass matrices does not change qualitatively the
results. There are essentially two conditions for the top quark to be
the heaviest fermion. The first is (for $g_1 = 0$)
\bea
tg^2 \beta > {2 M^2 + m_1^2 \over 2 M^2 + m_2^2 } \ . \label{eq:ar58}
\eea
where $m_1,m_2$ are the supersymmetric mass terms for the two Higgs.
The second is a rather involved lower bound for the dilaton vacuum
expectation value, so that  the underlying string theory must be in a
perturbative regime. We therefore need a minimal critical value for $tg
\beta$ of order one,  which depends on the soft masses, in order to have
a heavy top quark. Under these two assumptions, there is no need of fine
tuning to obtain a value of $\lambda$ of order $0.2$ which allows to
understand the hierarchy between the top quark and the other fermions.

 \section{Search for realistic orbifold examples.}
\vskip .5cm
We saw in Section 2 that a wide spread of modular weights is needed in order to
reproduce the
mass matrix hierarchies. In the case of orbifold compactifications,
modular weights are computable numbers for each given model. We consider
here the case of abelian symmetric orbifolds and evaluate which class
of models is selected by the requirement of the preceding section. We
will closely follow the approach of Ref. \cite{IL}.

The K\"ahler
potential of the matter fields has a simple dependence on the three
generic moduli $T_\alpha$ and on three model-dependent fields $U_m$. The
corresponding modular weights are denoted by $n^{(\alpha )}_i$ and
$l^{(\alpha)}_i$. \vskip .5cm The spectrum fields consists of two
sectors :
\vskip .5cm
$\bullet$ The untwisted sector, corresponding to
the string boundary conditions \linebreak $(i = 1 \cdots 6)$ \bea
X^i(\sigma = 2\pi , t) = X^i(\sigma = 0 , t) + V^i, \label{eq:ar36}
\eea
where $V^i$ are shifts in the six-dimensional lattice obtained by the action of
the space group.
\vskip .5cm
$\bullet$ The twisted sector, corresponding to the string boundary conditions
\bea
X^i(\sigma = 2\pi , t) = \theta X^i(\sigma = 0 , t) + V^i,
\label{eq:ar37}
\eea
where $\theta$ is a  twist which is an automorphism of the
six-dimensional lattice (some discrete rotation) $\theta^N = 1$, where
$N$ is called the order of the twist. \vskip .5cm
In order to have $N=1$ supersymmetry, $\theta$ must belong to $SU(3)$. The
twist $\theta$ can
be generally written as
\bea
\theta = \exp 2\pi i [v_1 J_{12} + v_2 J_{34} + v_3 J_{56} ],
\label{eq:ar38} \eea
where $N v_i \in \Z$ and the $J_{mn}$ are the $SO(6)$ Cartan
generators. $N=1$ supersymmetry implies $ \pm v_1 +  \pm v_2 + \pm v_3 =
0$ for some choice of signs.
\vskip .5cm
A generic twisted oscillator state has the form
\bea
\prod^3_{\beta, \gamma = 1} \prod_{m_\beta , n_\gamma } (\alpha^\beta_{m_\beta
+
\theta_\beta} )^{p^\beta_m}
(\widetilde{\alpha}^\gamma_{n_\gamma-\theta_\gamma} )^{q^\gamma_n} |
vacuum > , \label{eq:ar39} \eea
where $\alpha^\beta$ and $\widetilde \alpha^\gamma$ are the creation
operators corresponding respectively to the analytic and antianalytic
oscillators in the compactified directions. The indices $m_\beta$ and
$n_\gamma$ are the orders of the corresponding oscillators. Defining the
total number of oscillators as $p^\beta = \sum_m p^\beta_m$ and
$q^\gamma = \sum_n q^\gamma_n$ , the modular weights of the oscillator
states are, for the twisted sector,
\bea
&n^{(\alpha )}_i = - (1 - \theta^\alpha + p^\alpha_i - q^\alpha_i ) \ ,
l^{(\alpha)}_i = - (1 -
\theta^\alpha + q^\alpha_i - p^\alpha_i ) \; {\rm if} \; \theta^\alpha
\not= 0 \ , \nn \\ & \label{eq:ar40} \\
&n^{(\alpha )}_i = l^{(\alpha )}_i = 0 \ \; {\rm if} \; \theta^\alpha = 0
\ . \nn \eea
The states in the untwisted sector are characterized by
\bea
n^{(\alpha )}_\beta = - \delta^\alpha _\beta \ , l^{(\alpha )}_\beta = -
\delta^\alpha _\beta \ .
\label{eq:ar41}
\eea
The relevant quantity to evaluate the mass matrices is
$n^{(\alpha )}_i - n^{(\alpha )}_j$. The states $i$ and $j$ can be either
in the untwisted or in the twisted sector. We are searching for states
such that $n^{(\alpha )}_i - n^{(\alpha )} _j$ is maximum in order to
obtain the asymmetric relations eqs.(\ref{eq:ar20}). If, for example,
both states $i$ and $j$ are in the sector described by the same twist
vector, we obtain by using eq. (\ref{eq:ar40}) \bea n^{(\alpha )}_i -
n^{(\alpha )}_j = p^{(\alpha )}_j + q^{(\alpha )}_i - \big ( p^{(\alpha
)}_i + q^{(\alpha )}_j \big ) . \label{eq:ar42} \eea
\vskip .5cm
Since the low energy particles are to be found in the massless spectrum
of the string,  we are a priori interested only in the modular weights of
the massless particles. Some constraints can be obtained from the
mass formula for the left-moving twisted states
\bea
{1 \over 8} M^2_L = N_{osc} - h_{KM} + E_0 - 1 \label{eq:ar43}
\eea
where $N_{osc}$ is the fractional oscillator number. $E_0$ is the zero-point
 energy of the twist
$\theta$ given by the formula
\bea
E_0 = \sum^3_{\alpha=1} {1 \over 2} |v^\alpha | (1 - |v^\alpha|)  ,
\label{eq:ar44} \eea
where $v^\alpha$ is defined in eq.(\ref{eq:ar38}). The constant $h_{KM}$
is the contribution to the conformal dimension of the matter fields from
the left-moving $E_8 \times E_8$ gauge part. If the gauge group is $G =
\prod_a G_a$ and the massless particles are in the representations $R_a$
of $G_a$, then  \bea
h_{KM} = \sum_a {C(R_a) \over C(G_a) + k_a} \ , \label{eq:ar45}
\eea
where $k_a$ is the Kac-Moody level of the gauge factor $G_a$. In the following,
we consider a
model containing the spectrum of the MSSM and possibily extra matter or (and)
gauge
interactions. Therefore $h_{KM}$ computed within the MSSM gives a lower
bound to the real value. The minimum number of the oscillator states is
given by \bea
&p_{\max} &\leq \ N(1 - E_0 - h_{KM}) \ , \\
&q_{\max} &\leq \ {1 \over 1 - \theta^j}\  (1 - E_0 - h_{KM}) \ , \nn \\
 \label{eq:ar46}
\eea
where $N$ is the order of the twist $\overrightarrow \theta$.

Armed with these results we can readily check that the relations
(\ref{eq:ar20}) are impossible to
satisfy at the Kac-Moody levels $k_2 = k_3 = {3 \over 5} k_1 = 1$.

At  level two, a complete
scan of the abelian orbifolds gives the results displayed in Table 1.

The most difficult relation to satisfy, which is therefore our main
concern, is $n^{(\alpha )}_t - n^{(\alpha )}_u = 4$ , for some complex
plane $\alpha$. In table 1, only two models approximatively satisfy
it, the $\Z_{12}$ and $\Z_{12}'$ orbifolds with $(|v_1| , |v_2| , |v_3|
) = ({1 \over 3} , {1 \over 12} , {5 \over 12} )$  and $({1 \over 2} ,
{1 \over 12} , {5 \over 12} )$ respectively. The right-handed up quark
should be in the first or the fifth twisted sector $\theta (\theta^{12}
= 1 )$ and $(p,q) = (3 , 0 )$ with respect to the second or the third
complex  plane. The top  right-handed quark is
in the untwisted sector with $n^{(2)}_u = 0$. In this example $t_1 = t_3
= 1$ , $\varepsilon = t_2 / t_3 \sim (0.22)^2$. The second and the
third complex planes are
 completely rotated by the twist
vectors so the modular anomalies related to it must be cancelled completely by
the
Green-Schwarz mechanism.

More possibilities are allowed at Kac-Moody level three, as seen in Table 2.
The orbifolds which
can accommodate the hierarchy are
\vskip .5cm
$\bullet \quad \Z _8$ and $\Z_8'$ of twists $({1 \over 2} , {1 \over 8}
, {3 \over 8})$ and $({1 \over 4} , {1 \over 8} , {3 \over 8})$,
respectively. In both cases the right-handed up quark should be in the
first or the third twisted sector and $(p,q) = (3,0)$ with respect to
the second or the third complex plane. The right-handed top quark is in
the untwisted sector with $n^{(2)}_t = 0$. The anomalies with respect to
the second and the third planes are completely cancelled by the
Green-Schwarz mechanism.
\vskip .5cm
$\bullet \quad \Z _{12}$ and $\Z_{12}'$ of twists $({1 \over 3} , {1
\over 12} , {5 \over 12})$ and $({1 \over 2} , {1 \over 12} , {5 \over
12})$. The right-handed up quark must be in the first or the fifth
twisted sector, but more assignments for $(p,q)$ can be given. The
right-handed top  quark can be in the untwisted or in the twisted
sector. Another possibility for the up-quark is the second twisted
sector of $\Z_{12}'$ of twist  $(0, {1 \over 6}, {5 \over 6})$. The
anomalies with respect to the second  and the third planes are
completely cancelled by the Green-Schwarz mechanism.
\vskip .5cm
$\bullet \quad \Z_{2} \times \Z _{6}$ , $\Z_{3} \times \Z_{6}$ and
$\Z_{6} \times \Z_{6}$ of twists $(0, {1 \over 6} , {5 \over 6})$. The
example is similar to the $ \Z _8$, $\Z_8'$ cases, with the exception of
the anomalies.  Here all of the
three complex planes are left unrotated by a particular twist.
Consequently there are threshold corrections in the gauge coupling
constants which can partially cancel the modular anomalies. Moreover,
because of the fact that $p^{H_1,H_2}_{max} = 3$, the relation
(\ref{eq:ar134}) can be satisfied. so these models have the  possibility
of accomodating a phenomenologically correct unification scale $M_U$.
\vskip .5cm We also display in Table 3 an example for the case (iii) of
Section 2, which uses two small parameters $t_2 / t_1 \sim \lambda, \;
t_1 / t_3 \sim \lambda$. Only the oscillators for the right-handed
up-quarks are displayed, the others being easy to obtain.

As a general rule, the higher the Kac-Moody level, the simpler it is to
get mass hierarchies due to a wider spread of the allowed modular
weights. Such models have recently received attention in an
attempt of constructing grand unified string theories \cite{FIQ,AFIU}.

In general, the hierarchy appears as follows. For a modulus
corresponding to a small parameter, the second family fermions
should have more string oscillators compared to the third family and
the first family more than the second one (the opposite being true for a
modulus corresponding to a large parameter). The hierarchy thus
translates into a decreasing number of  allowed oscillators when going
from the light to the heavy families.

 \section{Concluding remarks.}

In this paper we analyzed the structure of the fermion mass matrices in
the effective superstring theories. It is found that, in some cases of
phenomenological interest, they are similar to the structures obtained
by imposing abelian horizontal symmetries. The analog of the abelian
charges are the modular weights of the matter fields; the small
expansion parameters are provided by the vev's of some moduli fields
away from their self-dual values. Hierarchical structures for the mass matrices
are obtained by assigning different modular weights for the three families
of quarks and leptons with respect to some moduli fields. A particular
case of interest is when the Yukawas are homogeneous functions of the
moduli, which can be viewed as a consequence of a 'diagonal' modular
symmetry of the theory, in the case where the original string couplings are
pure numbers. An interesting consequence is that the squark and
slepton mass matrices are proportional to the identity matrix. Consequently
they give no contributions to the FCNC processes like $b \rightarrow s \gamma$
or $\mu \rightarrow e \gamma$.

We stressed an intriguing connection between the mass matrices and the modular
anomalies, similar to the one between mass matrices and
mixed gauge anomalies in the horizontal symmetry approach recently
discussed in the literature. A phenomenologically relevant mass spectrum
requires one-loop modular anomalies, which can be cancelled in two ways.
The first one is the Green-Schwarz mechanism of superstrings. In this context,
if the Yukawa couplings are homogeneous functions of moduli and if the sum of
the modular
weights of the two Higgs doublets of the MSSM is symmetric in the moduli, then
a correct mass pattern asks for a Green-Schwarz mechanism with $k_1 = {5
\over 3}$ and the Weinberg angle is predicted to be $\sin^2 \theta_W
= {3 \over 8}$. The second way uses the moduli dependent threshold
corrections to the gauge coupling constants. In this case we obtain a relation
between the fermion masses, modular weights and the unification scale
$M_U$. Our analysis shows that we can acommodate a low value $M_U \sim M_s /
50$
provided the Higgs modular weights satisfy a constraint which is allowed
at Kac-Moody level two or three in abelian orbifolds. Hence we have the
possibility of a succesful unification scheme.

We have also investigated a dynamical mechanism for understanding the
fermion masses as a low-energy minimization process, previously restricted
to the top and bottom couplings. We show that the mechanism is easily
generalized to account
for the whole structure of the mass matrices, provided two inequalities on
the modular weights hold.

We have given orbifold examples where the hierarchies of the type that we
propose are
allowed. There are no examples at Kac-Moody level one due to the limited
range of the allowed modular weights, but we give examples at level two and
three.

There are, of course, many open questions and problems which deserve further
investigations. First of all the vev's of the moduli fields should be fixed
by the dynamics, which usually prefers the self-dual points. In the dynamical
approach, it would be also interesting to view the determination of the
Yukawa couplings directly from the point of view of the moduli fields:
in particular why the corresponding flat directions remain unlifted down to
low energies.

Finally it would be interesting to construct explicit orbifold models with
hierarchical mass matrices along these lines and to investigate their
phenomenological virtues.

\vskip 1.2cm
{\bf Acknowledgements}
\vskip .8cm
We would like to thank V.I. Zakharov for interesting discussions and
comments.

\newpage
{\bf Table 1.} Maximum number of allowed oscillators in abelian orbifolds for
\linebreak $(3/5) k_1 = k_2 = k_3 = 2$.
\begin{table}[h]
\centering
\begin{tabular} {|c|c|c|c|c|c|}
\hline
& & & & & \\
$ E_0$ &$|v_1 | , |v_2 | , |v_3 |$ &$ h^Q_{KM} = {37 \over 80} $ &$ h^U_{KM} =
{2 \over 5} $ &
$h^D_{KM} = {3 \over 10} $ &$ h^{H_1,H_2}_{KM} = {21 \over 80} $\\
& & & & & \\
\hline
& & & & & \\
$0$ &$(0,0,0)$ &$(p , q ) = (0,0)$ & $(0,0)$ &$(0,0)$ &$(0,0)$\\
& & & & & \\
${1 \over 3}$ &$ ({1 \over 3} , {1\over 3} , {2 \over 3} )$ & $(0,0)$ &
$(0,0) $ & $(1,0)$ &$(1,0)$\\ & & & & & \\
${5 \over 16} $ & $({1 \over 2} , {1 \over 4} , {1 \over 4} ) $ & $(0,0)$
&$(1,0)$ &$(1,0)$ &$(1,0)$\\ & & & & & \\
${1 \over 4}  $ & $({1 \over 3} , {1 \over 6} , {1 \over 6} ) $ & $(1,0)$
&$(2,1)$ &$(2,1)$ &$(2,1)$\\ & & & & & \\
${11 \over 36}  $ & $({1 \over 2} , {1 \over 3} , {1 \over 6} ) $ &
$(1,0)$ &$(1,0)$ &$(2,0)$ &$(2,0)$\\ & & & & & \\
${14 \over 49}  $ & $({3 \over 7} , {2 \over 7} , {1 \over 7} ) $ &
$(1,0)$ &$(2,0)$ &$(2,0)$ &$(3,1)$\\ & & & & & \\
${19 \over 64}  $ & $({1 \over 2} , {1 \over 8} , {3 \over 8} ) $ &
$(1,0)$ &$(2,0)$ &$(3,0)$ &$(3,0)$\\ & & & & & \\
${17 \over 64}  $ & $({1 \over 4} , {1 \over 8} , {3 \over 8} ) $ &
$(2,0)$ &$(2,0)$ &$(3,1)$ &$(3,1)$\\ & & & & & \\
${13 \over 48}  $ & $({1 \over 3} , {1 \over 12} , {5 \over 12} ) $ &
$(3,0)$ &$(3,0)$ &$(5,1)$ &$(5,1)$\\ & & & & & \\
${41 \over 144}  $ & $({1 \over 2} , {1 \over 12} , {5 \over 12} ) $ &
$(3,0)$ &$(3,0)$ &$(5,0)$ &$(5,0)$\\ & & & & & \\
${1 \over 4}  $ & $(0 ,{1 \over 2} , {1 \over 2} ) $ & $(0,0)$ &$(0,0)$
&$(0,0)$ &$(0,0)$\\ & & & & & \\
${2 \over 9}  $ & $(0 , {1 \over 3} , {1 \over 3} ) $ & $(0,0)$ &$(1,1)$
&$(1,1)$ &$(1,1)$\\ & & & & & \\
${3 \over 16}  $ & $(0 , {1 \over 4} , {1 \over 4} ) $ & $(1,1)$ &$(1,1)$
&$(2,2)$ &$(2,2)$\\ & & & & & \\
${5 \over 36}  $ & $(0 , {1 \over 6} , {1 \over 6} ) $ & $(2,2)$ &$(2,2)$
&$(3,3)$ &$(3,3)$\\ & & & & & \\
\hline
\end{tabular}
\end{table}

\newpage
{\bf Table 2.} Maximum number of allowed oscillators in abelian orbifolds
for\linebreak
 $(3/5) k_1 = k_2 = k_3 = 3$.
\begin{table}[h]
\centering
\begin{tabular} {|c|c|c|c|c|c|}
\hline
& & & & &  \\
$ E_0$ &$|v_1 | , |v_2 | , |v_3 |$ &$ h^Q_{KM} = {17 \over 45} $ &$ h^U_{KM} =
{14 \over 45}$
& $h^D_{KM} = {11 \over 45}$ &$h^{H_1,H_2}_{KM} = {1 \over 5} $\\
& & & & & \\
\hline
& & & & & \\
$0$ &$(0,0,0)$ &$(p , q ) = (0,0)$ & $(0,0)$ &$(0,0)$ &$(0,0)$\\
& & & & & \\
${1 \over 3}$ &$ ({1 \over 3} , {1\over 3} , {2\over 3} )$ & $(0,0)$ &
$(1,0) $ & $(1,0)$ &$(1,0)$\\ & & & &  &\\
${5 \over 16} $ & $({1 \over 2} , {1 \over 4} , {1\over 4} ) $ & $(1,0)$
&$(1,0)$ &$(1,0)$ &$(1,0)$\\ & & & & &\\
${1 \over 4}  $ & $({1 \over 3} , {1 \over 6} , {1\over 6} ) $ & $(2,1)$
&$(2,1)$ &$(3,1)$ &$(3,1)$\\ & & & & & \\
${11 \over 36}  $ & $({1 \over 2} , {1 \over 3} , {1\over 6} ) $ &
$(1,0)$ &$(2,0)$ &$(2,0)$ &$(2,0)$\\ & & & & & \\
${14 \over 49}  $ & $({3 \over 7} , {2 \over 7} , {1\over 7} ) $ &
$(2,0)$ &$(2,0)$ &$(3,1)$ &$(3,1)$\\ & & & & & \\
${19 \over 64}  $ & $({1 \over 2} , {1 \over 8} , {3\over 8} ) $ &
$(2,0)$ &$(3,0)$ &$(3,0)$ &$(4,1)$\\ & & & & &\\
${17 \over 64}  $ & $({1 \over 4} , {1 \over 8} , {3\over 8} ) $ &
$(2,0)$ &$(3,1)$ &$(3,1)$ &$(4,1)$\\ & & & & &\\
${13 \over 48}  $ & $({1 \over 3} , {1 \over 12} , {5\over 12} ) $ &
$(4,0)$ &$(5,1)$ &$(5,1)$ &$(6,1)$\\ & & & & & \\
${41 \over 144}  $ & $({1 \over 2} , {1 \over 12} , {5\over 12} ) $ &
$(4,0)$ &$(5,0)$ &$(5,0)$ &$(6,1)$\\ & & & & & \\
${1 \over 4}  $ & $(0 ,{1 \over 2} , {1\over 2} ) $ & $(0,0)$ &$(0,0)$
&$(1,1)$ &$(1,1)$\\ & & & & & \\
${2 \over 9}  $ & $(0 , {1 \over 3} , {\over 3} ) $ & $(1,1)$ &$(1,1)$
&$(1,1)$ &$(1,1)$\\ & & & &  &\\
${3 \over 16}  $ & $(0 , {1 \over 4} , {1\over 4} ) $ & $(1,1)$ &$(2,2)$
&$(2,2)$ &$(2,2)$\\ & & & & & \\
${5 \over 36}  $ & $(0 , {1 \over 6} , {1\over 6} ) $ & $(2,2)$ &$(3,3)$
&$(3,3)$ &$(3,3)$\\ & & & & & \\
\hline
\end{tabular}
\end{table}

\newpage
{\bf Table 3.} $\Z_{12}$ orbifold example for case (iii) with three
moduli and two small parameters, at Kac-Moody level three.
\begin{table}[h]
\centering
\begin{tabular} {|c|c|c|c|c|}
\hline
& & & &  \\
quark &twisted sector & $(p^{(1)},q^{(1)})$
&$(p^{(2)},q^{(2)})$ & $(p^{(3)},q^{(3)})$ \\ & & & &  \\
\hline
& & & &  \\
$u$ &$\theta$&$(0,0)$ & $(4,0)$ &$(0,0)$ \\
& & & &  \\
$c$ &$\theta^2$ & $(1,0)$ &
$(2,0) $ & $(1,0)$ \\ & & & &  \\
$t$ & $\theta^5 $ & $(0,0)$
&$(0,0)$ &$(4,0)$ \\ & & & &  \\
\hline
\end{tabular}
\end{table}

\end{document}